\documentclass[prl, superscriptaddress, showpacs, floatfix,
twocolumn]{revtex4}

\usepackage[]{graphicx}
\usepackage{amsmath}
\usepackage{amsfonts}

\def\one{{\mathchoice{\rm 1\mskip-4mu l}{\rm 1\mskip-4mu l}{\rm 1\mskip-4.5mu l}{\rm
1\mskip-5mu l}}}
\def\ket#1{| #1 \rangle}

\def\kb#1#2{|#1\rangle\!\langle #2 |}

\def\A{{\cal A}}
\def\B{{\cal B}}

\def\D{{\cal D}}
\def\E{{\cal E}}
\def\F{{\cal F}}
\def\H{{\cal H}}
\def\K{{\cal K}}

\newcommand{\fA}{{\mathfrak{A}}}

\def\Tr{{\mathrm{Tr}}}

\newcommand{\bbC}{{\mathbb{C}}}
\newtheorem{theorem}{Theorem}

\newtheorem{corollary}{Corollary}

\pacs{03.67.Pp, 03.67.Hk, 03.67.Lx}
\date{\today}

\begin{document}

\title{A Method To Find Quantum Noiseless Subsystems}

\author{Man-Duen~Choi}
\affiliation{Department of Mathematics, University of Toronto, ON
Canada, M5S 3G3}


\author{David~W.~Kribs}
\affiliation{Department of Mathematics and Statistics, University
of Guelph, Guelph, ON, Canada, N1G 2W1} \affiliation{Institute for
Quantum Computing, University of Waterloo, ON Canada, N2L 3G1}

\begin{abstract}
We develop a structure theory for {\it decoherence-free subspaces}
and {\it noiseless subsystems} that applies to arbitrary (not
necessarily unital) quantum operations. The theory can be
alternatively phrased in terms of the superoperator perspective,
or the algebraic noise commutant formalism. As an application, we
propose a method for finding all such subspaces and subsystems for
arbitrary quantum operations. We suggest that this work brings the
fundamental passive technique for error correction in quantum
computing an important step closer to practical realization.
\end{abstract}

\maketitle



\noindent{\it Introduction.} --- The problem of controlling and
maintaining properties of quantum systems which are in contact
with an environment has received considerable recent attention.
Primarily these investigations have been driven by the need to
better understand the special features of evolving quantum systems
that distinguish the quantum computing paradigm. Of central
importance in the field of ``quantum error correction'' is the
requirement for techniques to avoid and overcome the degrading
effects of decoherence. Early work in quantum error correction
included the realization that many physical error models contain
symmetries induced by the system-environment interplay. This led
to the discovery of ``decoherence-free subspaces'' (DFS) and
``noiseless subsystems'' (NS)
\cite{PSE96,DG97c,ZR97c,LCW98a,KLV00a,Zan01b,KBLW01a}. (On
occasion we shall refer to both notions jointly as ``NS''.) In
these schemes, subspaces -- and subsystems in the more abstract
case -- are identified within a system Hilbert space, with the
property that all initial states encoded therein remain immune to
the errors of a quantum operation of interest. Experimental
efforts \cite{KBAW00,KMRSIMW01,FVHTC02,VFPKLC03} have affirmed the
viability of this ``passive quantum error correction'' (PQEC)
technique.

It is also becoming clear that the NS formalism is applicable
beyond the realm of quantum error correction. In quantum
communication and cryptography, for instance, NS have been used as
vehicles for avoiding noise \cite{BGLPS04}; this may lead to
practical applications of NS in the near future. Further, NS are
ideal for determining how to achieve distributed quantum
information processing in the absence of shared reference frames
\cite{BRS03}. The NS concept has also arisen in recent analysis of
black holes \cite{DMS04}, and quantum gravity \cite{KM05}, where
NS are used to identify the relational symmetry-invariant physical
degrees of freedom in the quantum causal history framework.

There is an obvious advantage to PQEC in the context of quantum
computing. If a quantum operation (or channel) is found to possess
NS, then, by taking care at the initial encoding stage, the need
for active error correction after the fact is minimized. However,
this protocol has a notable drawback. While substantial analysis
has been carried out in important special cases, the protocol
lacks a general method to find NS for arbitrary quantum
operations. It is our belief that the PQEC approach will play a
substantive role in quantum computing devices, and in applications
beyond quantum computing, only if some sort of general approach
for finding NS is derived.

In this paper we propose such a method. Specifically, we develop a
structure theory that shows precisely how properties of a quantum
operation as a superoperator determine its NS structure. Moreover,
if an operator-sum decomposition of Kraus (or ``error'') operators
$\E = \{E_a\}$ for a channel $\E$ is known, we show how algebraic
properties of the operators $E_a$ determine this structure. This
information naturally leads to the aforementioned method. Our
analysis utilizes the framework for NS recently introduced under
the umbrella of ``operator quantum error correction'' in
\cite{KLP05,KLPL05} (see also \cite{Seife05}).

As a consequence of this work, we suggest that the fundamental
passive technique for error correction in quantum computing has
been brought an important step closer to practical realization.
Let us discuss these points further through a pair of illustrative
examples, full details of the theory will be provided below.


First we consider a simple example. Let $\H = \bbC^2\otimes\bbC^2$
be the combined system Hilbert space for two spin-$\frac 12$
particles. Let $\{\ket{00},\ket{01},\ket{10},\ket{11}\}$ be the
associated basis. Consider the channel $\E = \{Z_1,Z_2\}$ where
$Z_1 = Z\otimes\one_2$ and $Z_2 = \one_2\otimes Z$, with the Pauli
matrix $Z =\kb{0}{0}- \kb{1}{1}$. Then the action of $\E$ on a
density matrix $\rho$ on $\H$ is given by $\E(\rho) = \frac 12
(Z_1\rho Z_1^\dagger + Z_2\rho Z_2^\dagger)$. This channel has no
non-trivial NS. The key point is that the ``noise commutant''
$\A^\prime = \{Z_1,Z_2\}^\prime$, which is the set of all
operators on $\H$ that commute with both $Z_1$ and $Z_2$, only
contains the diagonal matrices with respect to the standard basis;
i.e., the matrices corresponding to classical states. On the other
hand, suppose our channel is $\F = \{U_1,U_2\}$, where $U_k =
UZ_k$, $k=1,2$, and $U = \one_4 - 2\,\kb{11}{11}$. In this case,
the noise commutant $\A^\prime = \{U_1,U_2\}^\prime$ contains a
single qubit NS (in fact it is a DFS). Indeed, any operator of the
form $\sigma = a \kb{00}{00} + b \kb{00}{11} +c \kb{00}{11} +d
\kb{11}{11}$, $a,b,c,d\in\bbC$, belongs to $\A^\prime$ and
satisfies $\E(\sigma)=\sigma$. As discussed below, that this NS is
also fixed by $\E$ follows from the fact that $\E$ is unital, or
bistochastic; i.e., $\E(\one)=\one$.

As a new example of NS, and one that will also be discussed
further below, we consider an error model first discussed in
\cite{KL97a} in the context of active error correction. In this
case the channel $\E = \{E_0,E_1,E_2\}$ acts on 2-qubit space and
has three Kraus operators given by
\begin{eqnarray}\label{nonunitaleg}
E_0 &=& \alpha (\kb{00}{00} + \kb{11}{11}) + \kb{01}{01} +
\kb{10}{10}, \quad
\\
E_1 &=& \beta (\kb{00}{00} + \kb{10}{00} + \kb{01}{11} +
\kb{11}{11}),
\\
E_2 &=& \beta (\kb{00}{00} - \kb{10}{00} - \kb{01}{11} +
\kb{11}{11}),
\end{eqnarray}
where $q$ is a scalar $0<q<1$ with $\alpha=\sqrt{1-2q}$ and
$\beta=\sqrt{q/2}$. One can check that $\E(\one)=\sum_{i=0}^2 E_i
E_i^\dagger \neq \one$, and hence $\E$ is non-unital. As we show
below, the noise commutant here $\A^\prime =
\{E_0,E_1,E_2\}^\prime$ supports a single qubit NS that is not
fixed by the action of $\E$. Further, there is another NS for the
channel, in fact a DFS, that is not contained in the noise
commutant. In particular, if we define the projector $P=
\kb{01}{01} + \kb{10}{10}$, then all operators supported by $P$
are fixed by $\E$; that is, $\E(\sigma)=\sigma$ for all $ \sigma =
P\sigma P$. However, these operators do not belong to the noise
commutant. For instance, notice that $E_iP=0\neq PE_i$ for
$i=1,2$. Hence, this error model has a NS inside its noise
commutant that is not fixed, and a fixed DFS that is not contained
in its noise commutant.

Thus, one can ask, what is the underlying phenomena that produces
noiseless subsystems?  The previous example indicates that we must
consider more than the noise commutant and fixed point set for the
map. As it turns out, the structure theory we derive for NS can be
phrased in terms of more general operator algebras obtained in the
same spirit as the noise commutant, and, alternatively, in terms
of modified fixed point sets for the map. Therefore, our approach
has the advantage of either being set in an algebraic context, or
strictly in terms of properties of the superoperator.

The rest of the paper is organized as follows. We next recall the
NS framework. We follow this by proving a theorem that yields the
structure theory, and then show precisely how it may be used to
find NS. Optimality of the method is then established, and this is
followed with a conclusion on possible future work and
limitations.


\noindent{\it Noiseless Subsystem Framework.} --- Given a quantum
operation (or ``channel''), represented by a completely positive,
trace preserving superoperator $\E:\B(\H)\rightarrow\B(\H)$ on a
(finite dimensional) Hilbert space $\H$, the NS protocol
\cite{PSE96,DG97c,ZR97c,LCW98a,KLV00a,Zan01b,KBLW01a,KLP05,KLPL05}
seeks subsystems $\H^B$ (with $\dim\H^B>1$) of the full system
Hilbert space $\H = (\H^A\otimes\H^B)\oplus\K$ such that
\begin{eqnarray}\label{nseqn}
\forall\sigma^A\ \forall\sigma^B,\ \exists \tau^A\ :\
\E(\sigma^A\otimes\sigma^B) = \tau^A\otimes\sigma^B.
\end{eqnarray}
Here we have written $\sigma^A$ (resp. $\sigma^B$) for operators
in $\B(\H^A)$ (resp. $\B(\H^B)$). In terms of partial traces,
Eq.~(\ref{nseqn}) can be equivalently phrased as,
\begin{eqnarray}\label{ptrace}
(\Tr_A \circ \E)(\sigma) = \Tr_A (\sigma), \quad \forall \sigma =
\sigma^A\otimes \sigma^B.
\end{eqnarray}
Thus, to be precise, $B$ is said to encode a {\it noiseless
subsystem} (or {\it decoherence-free subspace} in the case
$\dim\H^A=1$) for $\E:\B(\H)\rightarrow\B(\H)$ when
Eq.~(\ref{nseqn}) is satisfied.

The basic questions we address are the following: $(1)$ Is there a
structure theory for such subsystems?  $(2)$ If so, can it be
applied to derive a canonical method to find such subsystems for
arbitrary quantum operations? Our answer to the first question is
yes, and for the second we make a proposal that lends itself to
the possibility of a computational algorithm.





\noindent{\it Structure Theorem.} --- Let
$\E:\B(\H)\rightarrow\B(\H)$ be a quantum operation. We shall
write $\E=\{E_a\}$ when an error model for $\E$ is known; i.e.,
the operation elements $E_a$ determine $\E$ through the familiar
operator-sum representation $\E(\sigma)=\sum_a E_a\sigma
E_a^\dagger$ \cite{Cho75,Kra71}.

The (full) {\it noise commutant} $\A^\prime$ for $\E$ is the set
of all operators in $\B(\H)$ that commute with the operators $E_a$
and $E_a^\dagger$. The $\dagger$-algebra $\A$ generated by the
$E_a$ is called the {\it interaction algebra} associated with
$\E$. In the unital case ($\E(\one)=\one$) it is obvious that
every $\sigma\in\A'$ satisfies $\E(\sigma) = \sigma$, and, in
fact, every operator that is fixed by $\E$ belongs to $\A'$
\cite{Kri03a}. Of course, in the general case the operator
$\E(\one)$ may not be so well behaved, and all that can be said
for operators $\sigma\in\A'$ is that they satisfy $\E(\sigma) =
\sigma \E(\one) = \E(\one) \sigma$. This equation is suggestive of
the more general phenomena that must be analyzed to obtain NS for
arbitrary quantum operations. Given a projection $P$ in $\B(\H)$,
we shall make the natural  identification of the subalgebra
$P\B(\H)P$ of $\B(\H)$ with the algebra $\B(P\H)$.

\begin{theorem}\label{thm1}
Let $\E=\{E_a\}$ be a quantum operation on $\B(\H)$. Suppose $P$
is a projection on $\H$ such that
\begin{eqnarray}\label{nsprojns}
\E(P) = P\, \E(P) P.
\end{eqnarray}
Then $E_a P = P E_a P$, $\forall a$. Define
\begin{eqnarray*}\label{commid}
\A_P^\prime := \big\{ \sigma\in \B(P\H) : [\sigma,PE_a P] = 0 =
[\sigma, P E_a^\dagger P] \big\}.
\end{eqnarray*}
and,
\begin{eqnarray*}\label{modfixeqn}
{\rm Fix}_P(\E) := & &\big\{ \sigma\in \B(P\H): \E(\sigma) =
\sigma\E(P) = \E(P) \sigma, \\ & & \E(\sigma^\dagger \sigma) =
\sigma^\dagger \E(P) \sigma, \, \E(\sigma \sigma^\dagger) = \sigma
\E(P) \sigma^\dagger \big\},
\end{eqnarray*}
Then ${\rm Fix}_P(\E)$ is a $\dagger$-algebra inside $\B(P\H)$
that coincides with the algebra $\A_P^\prime$; that is,
\begin{eqnarray}
{\rm Fix}_P(\E) = \A_P^\prime.
\end{eqnarray}
\end{theorem}

{\noindent}{\it Proof.} Let $P$ be a projection that satisfies
Eq.~(\ref{nsprojns}). Then
\[
0 \leq P^\perp E_a P E_a^\dagger P^\perp \leq P^\perp \E(P)
P^\perp = 0 \quad \forall a.
\]
Hence $P^\perp E_a P =0$, or equivalently $E_aP=PE_aP$, $\forall
a$. Let $E_{a,P}:= PE_aP = E_a P$, $\forall a$. It is clear that
${\rm Fix}_P(\E)$ contains the commutant (taken inside $\B(P\H)$)
of the operators $\{E_{a,P}, E_{a,P}^\dagger \}$. Let $\sigma\in
\A^\prime_P$. We are required to show that $\sigma$ commutes with
the operators $E_{a,P}$ and $E_{a,P}^\dagger$.

The properties $\E(\sigma) = \sigma \E(P) = \E(P) \sigma$, $\sigma
= P\sigma P$ and Eq.~(\ref{nsprojns}) are seen through a
calculation to imply that
\[
\E(\sigma^\dagger \sigma) - \sigma^\dagger \E(P) \sigma = \sum_a
[\sigma,E_{a,P}^\dagger]^\dagger [\sigma,E_{a,P}^\dagger] \geq 0.
\]
(This inequality may be regarded as a generalization  of the
Schwarz inequality for completely positive maps from
\cite{Cho74,Dav57}.) Thus, given $\E(\sigma) = \sigma \E(P) =
\E(P) \sigma$, and so $\E(\sigma^\dagger) = \E(\sigma)^\dagger =
\sigma^\dagger \E(P) = \E(P) \sigma^\dagger$, it follows that
\begin{eqnarray*}
\E(\sigma^\dagger \sigma) = \sigma^\dagger \sigma \E(P) \,\,\,\,
&{\rm iff}& \,\,\,\, \sigma E_{a,P}^\dagger = E_{a,P}^\dagger
\sigma, \,\,\, \forall a, \\
\E(\sigma \sigma^\dagger) = \sigma \sigma^\dagger \E(P) \,\,\,\,
&{\rm iff}& \,\,\,\, \sigma E_{a,P} = E_{a,P} \sigma, \,\,\,
\forall a.
\end{eqnarray*}
This completes the proof.  \hfill$\square$

Observe that the maximally mixed state $P=\one$ trivially
satisfies Eq.~(\ref{nsprojns}), and the algebra $\A_\one^\prime$
coincides with the full noise commutant $\{E_a,E_a^\dagger\}$.
However, as discussed above, the operator $\E(\one)$ may not have
many nice properties. In general there may be other projections
$P$ that support larger noiseless subsystems.








\noindent{\it Noiseless Subsystems.} --- Let $P$ be a projection
that satisfies Eq.~(\ref{nsprojns}). The structure theory for
$\dagger$-algebras \cite{Dav96a} yields a unitary $U$ on $P\H$
such that
\begin{eqnarray}\label{ue}
U\, \A^\prime_P \, U^\dagger = \bigoplus_k (\one_{m_k}\otimes
M_{n_k}),
\end{eqnarray}
for a unique (up to reordering) family of positive integers
$m_k,n_k\geq 1$. We have used $M_{n_k}$ to denote the operator
algebra $\B(\bbC^{n_k})$, represented as matrices with respect to
some orthonormal basis. Note that the algebra $\A_P^\prime$ may be
regarded as a subalgebra of $\B(\H)$ simply by taking a direct sum
$\A^\prime_P\oplus O_{n_P}$ of $\A_P^\prime$ together with the
``zero algebra'' $0_{n_P}$ of $n_P\times n_P$ matrices on $P^\perp
\H$, where $n_P = {\rm dim}\,\H - \sum_k m_kn_k$.

The algebra structure Eq.~(\ref{ue}) induces a decomposition of
the subspace $P\H$ as
\begin{eqnarray}\label{spatial}
P\H = \bigoplus_k (\H^{A_k}\otimes\H^{B_k}),
\end{eqnarray}
where ${\rm dim}\,(\H^{A_k})=m_k$ and ${\rm dim}\,(\H^{B_k})=n_k$.
Observe that the positive operator $\E(P)$ belongs to the
commutant inside $\B(P\H)$ of $\A_P^\prime$ by definition. As this
commutant has structure $\A_P:= \A_P^{\prime\prime} = \oplus_k
(M_{m_k}\otimes\one_{n_k})$, it follows that there are operators
$\sigma_k\in \B(\H^{A_k})=M_{m_k}$ such that $\E(P) = \sum_k
\sigma_k\otimes\one_{n_k}$.

Now let $\rho = \one^{A_k}\otimes\rho^{B_k}$ belong to the
subalgebra $\one^{A_k}\otimes\B(\H^{B_k})$ of $\A_P^\prime
=\oplus_k (\one^{A_k}\otimes \B(\H^{B_k}))$. Then we have
\begin{eqnarray}\label{newns}
 \E(\one^{A_k}\otimes\rho^{B_k})\,=\,\E(\rho)\, =\, \rho\, \E(P) \,=
\,\sigma_k \otimes \rho^{B_k}.
\end{eqnarray}
But Eq.~(\ref{nseqn}) holds if and only if  it holds for $\sigma^A
= \one^A$ \cite{KLP05,KLPL05}. Therefore, it follows from
Eq.~(\ref{newns}) that each of the subsystems $\H^{B_k}$ is
noiseless for $\E$ and the following result is established.

\begin{theorem}
Let $\E$ be a quantum operation on $\B(\H)$. Let $P$ be a
projection on $\H$ that satisfies Eq.~(\ref{nsprojns}) and let
$P\H = \bigoplus_k (\H^{A_k}\otimes\H^{B_k})$ be the decomposition
of $P\H$ induced by the $\dagger$-algebra structure of
$\A_P^\prime = {\rm Fix}_P(\E)$. Then the subsystems $\H^{B_k}$,
with $\dim\H^{B_k}>1$, are each noiseless subsystems for $\E$.
\end{theorem}

In fact, it follows that if the input states are restricted to the
subspace $\H^{A_k}\otimes\H^{B_k}$, then the corresponding
restriction of $\E$ satisfies $\E(P_k(\cdot)P_k) = \E_k
\,\otimes\, {\rm id}_{B_k}$ where $P_k$ is the projection of $\H$
onto $\H^{A_k}\otimes\H^{B_k}$, $\E_k$ is a quantum operation on
$\B(\H^{A_k})$, and ${\rm id}_{B_k}$ is the identity channel on
$\B(\H^{B_k})$.

The NS structure for a number of unital channels have been
analyzed in detail. An extensively studied class of channels arise
from ``collective noise'', which has a number of physical
interpretations (see \cite{HKLP05,JKK05} for a detailed analysis
of this and related NS structures). We note a connection with
\cite{BRS04} which includes a decomposition for collective noise
channels of the form $\E = \sum_k (\D_k\otimes\,{\rm
id}_k)(P_k\sigma P_k)$, where the $P_k$ are projections associated
with a decomposition of the system Hilbert space induced by
underlying representation theory and the $\D_k$ are depolarizing
channels (see Eq.~(22) of \cite{BRS04}). Interestingly, this may
now be seen as a special case of the general form derived here.

Let us return to the non-unital example discussed in the
Introduction. A computation shows in this case that the full noise
commutant satisfies $\A^\prime_\one = \{E_0,E_1,E_2\}^\prime \cong
\one_2\otimes M_2$, and thus supports a single qubit NS. Indeed,
if $\sigma\in\A^\prime_\one$ is written as
$\one_2\otimes\sigma_0$, $\sigma_0\in M_2$, with respect to this
unitary equivalence, then a calculation shows that
\[
\E(\sigma) = \E(\one_2\otimes\sigma_0) = \left(
\begin{smallmatrix} 1 - q & 0 \\ 0 & 1+q \end{smallmatrix}\right)
\otimes \sigma_0.
\]

But recall that the projection $P= \kb{01}{01} + \kb{10}{10}$
defines a DFS for $\E$; specifically, $\E(\sigma)=\sigma$,
$\forall \sigma = P\sigma P$. Now we can see precisely how this
DFS arises. Namely, $\E(P)=P$ satisfies Eq.~(\ref{nsprojns}) and
thus we find a single qubit NS for $\E$, with
$\ket{\psi}:=\ket{01}$ and $\ket{\phi} := \ket{10}$, given by
\begin{eqnarray*}
\A_P^\prime = {\rm span}\,\big\{ \kb{\psi}{\psi}, \kb{\psi}{\phi},
\kb{\phi}{\psi}, \kb{\phi}{\phi} \big\}  \cong M_2.
\end{eqnarray*}

Notice that $\A_P^\prime$ is not contained in $\A_\one^\prime$,
and thus this DFS would not be detected through an analysis of the
full noise commutant alone. Further, while $\A^\prime_\one$ and
$\A^\prime_P$ have the same ``size'' from an encoding viewpoint
(i.e., a single qubit) \cite{Kup03}, it is perhaps more convenient
to work with $\A_P^\prime \cong M_2$, as it can be more easily
isolated within the full system Hilbert space.

In fact, this example gives an indication as to how active and
passive techniques for quantum error correction can be combined to
combat noise. Indeed, we have just noted that the subspace
$\{\ket{01},\ket{10}\}$ determines a DFS for $\E$. On the other
hand, in \cite{KL97a} it was shown that active error correction
may be used to overcome corruption by $\E$ of the code subspace
$\{\ket{00},\ket{11}\}$.


\noindent{\it Optimality of the Method.} --- The previous two
sections yield a canonical method to compute noiseless subsystems
for a given quantum operation $\E$ which can be succinctly stated
as follows:
\begin{itemize}
\item[(i)] Compute the projections $P$ such that
Eq.~(\ref{nsprojns}) holds. \item[(ii)] Compute the structure of
the algebras $\A_P^\prime={\rm Fix}_P(\E)$ as in Eq.~(\ref{ue}).
\end{itemize}
Then, in the notation above, the subspaces $\H^{B_k}$, with $\dim
\H^{B_k}>1$, encode noiseless subsystems for $\E$ via the operator
algebras $\one^{A_k}\otimes\B(\H^{B_k})$

A crucial final step in the process is to determine if this scheme
captures all noiseless subsystems for $\E$.  We next show that
this is indeed the case.

\begin{theorem}\label{inversethm}
Let $\E$ be a quantum operation on $\B(\H)$. Suppose that $\H =
(\H^A\otimes\H^B)\oplus\K$ and that $\H^B$ is a noiseless
subsystem for $\E$ as in Eq.~(\ref{nseqn}). Let $P$ be the
projection of $\H$ onto $\H^A\otimes\H^B$. Then $\E(P) = P\E(P) P$
and the algebra $\A_P^\prime$ contains  $\one^A \otimes \B(\H^B)$
as a simple, unital $\dagger$-subalgebra.
\end{theorem}

{\noindent}{\it Proof.} Let $\{\ket{\alpha_k}\}$ be an orthonormal
basis for $\H^A$, and let $ \{P_{kl} =
\kb{\alpha_k}{\alpha_l}\otimes\one^B\} $ be the corresponding
matrix units inside $\B(\H^A)\otimes\one^B$. It was proved in
\cite{KLP05,KLPL05} that $\H^B$ is noiseless for $\E=\{E_a\}$ as
in Eq.~(\ref{nseqn}) precisely when $E_aP=PE_aP$ and there are
scalars $\{\lambda_{akl}\}$ such that
\begin{eqnarray}\label{nscond}
P_{kk} E_a P_{ll} = \lambda_{akl} P_{kl} \quad \forall\, a,k,l.
\end{eqnarray}

Note that the projection $P$ is given by $P=\sum_k P_{k}$, where
we have written $P_k$ for $P_{kk}$. Thus we have
\begin{eqnarray*}
\E(P) & =& \sum_{a,k} E_a P_k E_a^\dagger =\sum_{a,k,l,l'} P_l E_a
P_k E_a^\dagger P_{l'} \\ &=& \sum_{a,k,l,l'}
\lambda_{alk}\overline{\lambda}_{al'k} P_{lk}P_{kl'} =
\sum_{a,k,l,l'} \lambda_{alk}\overline{\lambda}_{al'k} P_{ll'}.
\end{eqnarray*}
Let  $\sigma = \one^A\otimes\sigma^B\in\one^A\otimes\B(\H^B)$.
Then since the $P_{kl}$ commute with $\sigma=P\sigma P$ we have
\begin{eqnarray*}
\E(\sigma) &=& \E(P\sigma P) = P\E(P\sigma P) P  \\  &=&
\sum_{a,k,k',l,l'} P_{k}E_a P_{k'}\sigma P_{l'} E_a^\dagger P_{l}
\\ &=& \sum_{a,k,k',l,l'} \lambda_{akk'} \overline{\lambda}_{all'}
P_{kk'}\sigma P_{l'l} \\ &=& \sigma \E(P) = \E(P) \sigma.
\end{eqnarray*}

In particular, this implies (with $\sigma^B = \one^B$) that $\E(P)
= P\E(P)P$ and that the algebra $\one^A\otimes\B(\H^B)$ is
contained in $\A_P^\prime$. It is clear that $\A_P^\prime$ and
$\one^A\otimes\B(\H^B)$ have the same unit $P$, and that
$\one^A\otimes\B(\H^B)$ is a simple (i.e., contains no non-trivial
ideals) $\dagger$-subalgebra of $\A_P^\prime$. \hfill$\square$

We finish with a consequence for the unital case. The class of
unital channels includes numerous physical error models such as
collective noise, randomized unitary channels, etc. It is
important to note that the full noise commutant captures all NS in
this case. In particular, this means algebras $\A^\prime_P$ may
not be contained inside $\A^\prime_\one$ only in the non-unital
case.


\begin{corollary}
Let $\E$ be a unital quantum operation on $\B(\H)$. If $\fA
=\one^A\otimes\B(\H^B)$ is the algebra determined by a noiseless
subsystem for $\E$ as in Eq.~(\ref{nseqn}), then $\fA$ is a
subalgebra of the full noise commutant $\A^\prime_\one$.
\end{corollary}

{\noindent}{\it Proof.} Let $P$ be the projection of $\H$ onto
$\H^A\otimes\H^B$. By Eq.~(\ref{nseqn}), there is a $\tau^A$ such
that $\E(P) = \E(\one^A\otimes\one^B) = \tau^A\otimes\one^B$.
Since $\E$ is a unital completely positive map, we know that
$\tau^A$ is a contraction operator, and hence $\E(P) \leq P$. Then
in fact $\E(P)=P$ by Lemma~2.3 from \cite{Kri03a}. Thus, it
follows from Theorem~\ref{inversethm}, and the definition of
$\A_P^\prime$, that $\fA \subseteq \A_P^\prime$ is a subalgebra of
${\rm Fix}\,(\E)$.
 \hfill$\square$


\noindent{\it Conclusion.} --- We have derived a structure theory
for decoherence-free subspaces and noiseless subsystems that
applies to arbitrary quantum operations. As an application, we
have proposed a method to compute NS for any given operation. We
expect that the method could be formalized into a computational
algorithm, as suggested by recent literature \cite{HKL04,Zar05}
which includes algorithms written to calculate operator algebra
structures, but there are still details to work through. We plan
to undertake this investigation elsewhere.

We discussed a non-unital example in which the maximally mixed
state and a smaller projection support different single qubit
noiseless subsystems. We suggest that this work motivates
reconsideration of the quantum channels that appear in the
literature, for the possible existence of noiseless subsystems. We
wonder about possible experimental implications of this work. It
would also be interesting to investigate connections with other
recent noiseless subsystem related efforts such as
\cite{DMS04,LS05,KM05}.


\noindent{\it Acknowledgements.} We thank Dietmar Bisch for asking
a question that partly motivated this work. We are grateful to
John Holbrook, Raymond Laflamme, Rob Spekkens, and Karol
Zyczkowski for helpful comments. D.W.K. would also like to thank
other colleagues at UofG, IQC and Perimeter Institute for
interesting discussions. This work was partially supported by
NSERC.


\begin{thebibliography}{30}
\expandafter\ifx\csname
natexlab\endcsname\relax\def\natexlab#1{#1}\fi
\expandafter\ifx\csname bibnamefont\endcsname\relax
  \def\bibnamefont#1{#1}\fi
\expandafter\ifx\csname bibfnamefont\endcsname\relax
  \def\bibfnamefont#1{#1}\fi
\expandafter\ifx\csname citenamefont\endcsname\relax
  \def\citenamefont#1{#1}\fi
\expandafter\ifx\csname url\endcsname\relax
  \def\url#1{\texttt{#1}}\fi
\expandafter\ifx\csname
urlprefix\endcsname\relax\def\urlprefix{URL }\fi
\providecommand{\bibinfo}[2]{#2}
\providecommand{\eprint}[2][]{\url{#2}}






\bibitem[{\citenamefont{Palma et~al.}(1996)}]{PSE96}
\bibinfo{author}{\bibfnamefont{G.M.}~\bibnamefont{Palma}},
\bibinfo{author}{\bibfnamefont{K.-A.}~\bibnamefont{Suominen}},
  \bibinfo{author}{\bibfnamefont{A.}~\bibnamefont{Ekert}},
  \bibinfo{journal}{Proc. Royal Soc. A} \textbf{\bibinfo{volume}{452}},
  \bibinfo{pages}{567} (\bibinfo{year}{1996}).

\bibitem[{\citenamefont{Duan and Guo}(1997)}]{DG97c}
\bibinfo{author}{\bibfnamefont{L.-M.} \bibnamefont{Duan}},
  \bibinfo{author}{\bibfnamefont{G.-C.} \bibnamefont{Guo}},
  \bibinfo{journal}{Phys. Rev. Lett.} \textbf{\bibinfo{volume}{79}},
  \bibinfo{pages}{1953} (\bibinfo{year}{1997}).

\bibitem[{\citenamefont{Zanardi and Rasetti}(1997)}]{ZR97c}
\bibinfo{author}{\bibfnamefont{P.}~\bibnamefont{Zanardi}},
  \bibinfo{author}{\bibfnamefont{M.}~\bibnamefont{Rasetti}},
  \bibinfo{journal}{Phys. Rev. Lett.} \textbf{\bibinfo{volume}{79}},
  \bibinfo{pages}{3306} (\bibinfo{year}{1997}).

\bibitem[{\citenamefont{Lidar et~al.}(1998)\citenamefont{Lidar, Chuang, and
  Whaley}}]{LCW98a}
\bibinfo{author}{\bibfnamefont{D.A.}~\bibnamefont{Lidar}},
  \bibinfo{author}{\bibfnamefont{I.L.}~\bibnamefont{Chuang}},
  \bibinfo{author}{\bibfnamefont{K.B.}~\bibnamefont{Whaley}},
  \bibinfo{journal}{Phys. Rev. Lett.} \textbf{\bibinfo{volume}{81}},
  \bibinfo{pages}{2594} (\bibinfo{year}{1998}).

\bibitem[{\citenamefont{Knill et~al.}(2000)\citenamefont{Knill, Laflamme, and
  Viola}}]{KLV00a}
\bibinfo{author}{\bibfnamefont{E.}~\bibnamefont{Knill}},
  \bibinfo{author}{\bibfnamefont{R.}~\bibnamefont{Laflamme}},
  \bibinfo{author}{\bibfnamefont{L.}~\bibnamefont{Viola}},
  \bibinfo{journal}{Phys. Rev. Lett.} \textbf{\bibinfo{volume}{84}},
  \bibinfo{pages}{2525} (\bibinfo{year}{2000}).

\bibitem[{\citenamefont{Zanardi}(2001)}]{Zan01b}
\bibinfo{author}{\bibfnamefont{P.}~\bibnamefont{Zanardi}},
  \bibinfo{journal}{Phys. Rev. A} \textbf{\bibinfo{volume}{63}},
  \bibinfo{pages}{12301} (\bibinfo{year}{2001}).

\bibitem[{\citenamefont{Kempe et~al.}(2001)\citenamefont{Kempe, Bacon, Lidar,
  and Whaley}}]{KBLW01a}
\bibinfo{author}{\bibfnamefont{J.}~\bibnamefont{Kempe}},
  \bibinfo{author}{\bibfnamefont{D.}~\bibnamefont{Bacon}},
  \bibinfo{author}{\bibfnamefont{D.~A.} \bibnamefont{Lidar}},
  \bibinfo{author}{\bibfnamefont{K.~B.} \bibnamefont{Whaley}},
  \bibinfo{journal}{Phys. Rev. A} \textbf{\bibinfo{volume}{63}},
  \bibinfo{pages}{42307} (\bibinfo{year}{2001}).

\bibitem[{\citenamefont{Kwiat et~al.}(2000)\citenamefont{Kwiat, Berglund, Altepeter, White}}]{KBAW00}
\bibinfo{author}{\bibfnamefont{P.~G.}~\bibnamefont{Kwiat, et~al.}},
  \bibinfo{journal}{Science} \textbf{\bibinfo{volume}{290}},
  \bibinfo{pages}{498} (\bibinfo{year}{2000}).

\bibitem[{\citenamefont{Kielpinski et~al.}(2001)\citenamefont{Kielpinski, Meyer, Rowe, Sackett, Itano, Monroe, Wineland}}]{KMRSIMW01}
\bibinfo{author}{\bibfnamefont{D.}~\bibnamefont{Kielpinski, et~al.}},
  \bibinfo{journal}{Science} \textbf{\bibinfo{volume}{291}},
  \bibinfo{pages}{1013} (\bibinfo{year}{2001}).

\bibitem[{\citenamefont{Fortunato et~al.}(2002)\citenamefont{Fortunato, Viola, Hodges, Teklemariam, Cory}}]{FVHTC02}
\bibinfo{author}{\bibfnamefont{E.~M.}~\bibnamefont{Fortunato, et~al.}},
  \bibinfo{journal}{New J. Phys.} \textbf{\bibinfo{volume}{4}},
  \bibinfo{pages}{5} (\bibinfo{year}{2002}).

\bibitem[{\citenamefont{Viola et~al.}(2003)\citenamefont{Viola, Fortunato, Pravia, Knill, Laflamme, Cory}}]{VFPKLC03}
  \bibinfo{author}{\bibfnamefont{L.}~\bibnamefont{Viola, et~al.}},
  \bibinfo{journal}{Science} \textbf{\bibinfo{volume}{293}},
  \bibinfo{pages}{2059} (\bibinfo{year}{2003}).

\bibitem[{\citenamefont{Boileau et~al.}(2003)\citenamefont{Boileau, Gottesman, Laflamme, Poulin, Spekkens}}]{BGLPS04}
  \bibinfo{author}{\bibfnamefont{J.~-C.}~\bibnamefont{Boileau, et~al.}},
  \bibinfo{journal}{Phys. Rev. Lett.} \textbf{\bibinfo{volume}{92}},
  \bibinfo{pages}{17901} (\bibinfo{year}{2004}).

\bibitem{BRS03}
\bibinfo{author}{\bibfnamefont{S.~D.}~\bibnamefont{Bartlett}},
  \bibinfo{author}{\bibfnamefont{T.}~\bibnamefont{Rudolph}},
  \bibinfo{author}{\bibfnamefont{R.~W.}~\bibnamefont{Spekkens}},
  \bibinfo{journal}{Phys. Rev. Lett.},\textbf{\bibinfo{volume}{91}},
  \bibinfo{pages}{027901} (\bibinfo{year}{2003}).

\bibitem{DMS04}
\bibinfo{author}{\bibfnamefont{O.}~\bibnamefont{Dreyer}},
\bibinfo{author}{\bibfnamefont{F.}~\bibnamefont{Markopoulou}},
  \bibinfo{author}{\bibfnamefont{L.}~\bibnamefont{Smolin}},
  \bibinfo{journal}{arxiv.org/hep-th/0409056}.

\bibitem{KM05}
\bibinfo{author}{\bibfnamefont{D.~W.}~\bibnamefont{Kribs}},
\bibinfo{author}{\bibfnamefont{F.}~\bibnamefont{Markopoulou}},
  \bibinfo{journal}{arxiv.org/gr-qc/0510052}.

\bibitem{KLP05}
\bibinfo{author}{\bibfnamefont{D.~W.}~\bibnamefont{Kribs}},
  \bibinfo{author}{\bibfnamefont{R.}~\bibnamefont{Laflamme}},
  \bibinfo{author}{\bibfnamefont{D.}~\bibnamefont{Poulin}},
  \bibinfo{journal}{Phys. Rev. Lett.},\textbf{\bibinfo{volume}{94}},
  \bibinfo{pages}{180501} (\bibinfo{year}{2005}).

\bibitem{KLPL05}
\bibinfo{author}{\bibfnamefont{D.~W.}~\bibnamefont{Kribs}},
  \bibinfo{author}{\bibfnamefont{R.}~\bibnamefont{Laflamme}},
  \bibinfo{author}{\bibfnamefont{D.}~\bibnamefont{Poulin}},
  \bibinfo{author}{\bibfnamefont{M.}~\bibnamefont{Lesosky}},
 \bibinfo{journal}{arxiv.org/quant-ph/0504189}.

\bibitem[{\citenamefont{Seife}(2005)}]{Seife05}
\bibinfo{author}{\bibfnamefont{C.} \bibnamefont{Seife}},
 \emph{\bibinfo{title}{Teaching qubits new tricks}},  \bibinfo{journal}{Science}, \textbf{\bibinfo{volume}{309}},
  \bibinfo{pages}{238} (\bibinfo{year}{2005}).




\bibitem[{\citenamefont{Knill and Laflamme}(1997)}]{KL97a}
\bibinfo{author}{\bibfnamefont{E.}~\bibnamefont{Knill}} \bibnamefont{and}
  \bibinfo{author}{\bibfnamefont{R.}~\bibnamefont{Laflamme}},
  \bibinfo{journal}{Phys. Rev. {A}} \textbf{\bibinfo{volume}{55}},
  \bibinfo{pages}{900} (\bibinfo{year}{1997}).

\bibitem[{\citenamefont{Choi}(1975)}]{Cho75}
\bibinfo{author}{\bibfnamefont{M.~D.} \bibnamefont{Choi}},
  \bibinfo{journal}{Lin. Alg. Appl.} \textbf{\bibinfo{volume}{10}}
 \bibinfo{pages}{285-290} (\bibinfo{year}{1975}).

\bibitem[{\citenamefont{Kraus}(1971)}]{Kra71}
\bibinfo{author}{\bibfnamefont{K.} \bibnamefont{Kraus}},
  \bibinfo{journal}{Ann. Physics} \textbf{\bibinfo{volume}{64}}
 \bibinfo{pages}{311-335} (\bibinfo{year}{1971}).

\bibitem[{\citenamefont{Kribs}(2003)}]{Kri03a}
\bibinfo{author}{\bibfnamefont{D.~W.} \bibnamefont{Kribs}},
  \bibinfo{journal}{Proc. Edin. Math. Soc.} \textbf{\bibinfo{volume}{46}}
   \bibinfo{pages}{421-433}(\bibinfo{year}{2003}).

\bibitem[{\citenamefont{Davidson}(1996)}]{Dav96a}
\bibinfo{author}{\bibfnamefont{K.}~\bibnamefont{Davidson}},
  \emph{\bibinfo{title}{$\mathrm{C}^*$-algebras by example, Fields Institute
  Monographs}} (\bibinfo{publisher}{Amer. Math. Soc., Providence},
  \bibinfo{year}{1996}).

\bibitem[{\citenamefont{Choi}(1974)}]{Cho74}
\bibinfo{author}{\bibfnamefont{M.~D.} \bibnamefont{Choi}},
  \bibinfo{journal}{Illinois J. Math.} \textbf{\bibinfo{volume}{18}}
 \bibinfo{pages}{565-574} (\bibinfo{year}{1974}).

\bibitem[{\citenamefont{Davis}(1957)}]{Dav57}
\bibinfo{author}{\bibfnamefont{C.} \bibnamefont{Davis}},
  \bibinfo{journal}{Proc. Amer. Math. Soc.} \textbf{\bibinfo{volume}{8}}
 \bibinfo{pages}{42-44} (\bibinfo{year}{1957}).

\bibitem[{\citenamefont{Holbrook, et~al.}(2005)\citenamefont{Holbrook, Kribs, Laflamme, and Poulin}}]{HKLP05}
\bibinfo{author}{\bibfnamefont{J.~A.} \bibnamefont{Holbrook}},
  \bibinfo{author}{\bibfnamefont{D.~W.} \bibnamefont{Kribs}},
  \bibinfo{author}{\bibfnamefont{R.} \bibnamefont{Laflamme}},
  \bibinfo{author}{\bibfnamefont{D.}
  \bibnamefont{Poulin}}, \bibinfo{journal}{Integral Eqnts. \& Operator Thy.}
  \textbf{\bibinfo{volume}{51}}, \bibinfo{pages}{215-234} (\bibinfo{year}{2005}).

\bibitem[{\citenamefont{Junge et~al.}(2005)\citenamefont{Junge, Kribs, and Kim}}]{JKK05}
\bibinfo{author}{\bibfnamefont{M.} \bibnamefont{Junge}},
  \bibinfo{author}{\bibfnamefont{P.} \bibnamefont{Kim}},
  \bibinfo{author}{\bibfnamefont{D.~W.} \bibnamefont{Kribs}},
   \bibinfo{journal}{J. Math. Phys.}
  \textbf{\bibinfo{volume}{46}}, \bibinfo{pages}{022102} (\bibinfo{year}{2005}).

\bibitem[{\citenamefont{Bartlett et~al.}(2004)\citenamefont{Bartlett, Rudolph, and Spekkens}}]{BRS04}
\bibinfo{author}{\bibfnamefont{S.~D.} \bibnamefont{Bartlett}},
  \bibinfo{author}{\bibfnamefont{T.} \bibnamefont{Rudolph}},
  \bibinfo{author}{\bibfnamefont{R.~W.} \bibnamefont{Spekkens}},
   \bibinfo{journal}{Phys. Rev. A}
  \textbf{\bibinfo{volume}{70}}, \bibinfo{pages}{032307} (\bibinfo{year}{2004}).

\bibitem{Kup03}
\bibinfo{author}{\bibfnamefont{G.}~\bibnamefont{Kuperberg}},
  \bibinfo{journal}{IEEE Trans. Inform. Theory}  \textbf{\bibinfo{volume}{49}}, \bibinfo{pages}{1465-1473} (\bibinfo{year}{2003}).


\bibitem[{\citenamefont{Holbrook, et al.}(2004)\citenamefont{Holbrook, Kribs, and Laflamme}}]{HKL04}
\bibinfo{author}{\bibfnamefont{J.~A.}~\bibnamefont{Holbrook}},
  \bibinfo{author}{\bibfnamefont{D.~W.}~\bibnamefont{Kribs}},
  \bibinfo{author}{\bibfnamefont{R.}~\bibnamefont{Laflamme}},
  \bibinfo{journal}{Quantum Inf. Proc.} \textbf{\bibinfo{volume}{2}},
  \bibinfo{pages}{381} (\bibinfo{year}{2004}).

\bibitem[{\citenamefont{Zarikian}(2005)}]{Zar05}
\bibinfo{author}{\bibfnamefont{V.} \bibnamefont{Zarikian}},
  \bibinfo{journal}{Lin. Alg. Appl., to appear} .















\bibitem{LS05}
\bibinfo{author}{\bibfnamefont{D.~A.}~\bibnamefont{Lidar}}, 
  \bibinfo{author}{\bibfnamefont{A.}~\bibnamefont{Shabani}},
  \bibinfo{journal}{arxiv.org/quant-ph/0505051}.




\end{thebibliography}


\end{document}